\documentclass[twocolumn,aps,pre]{revtex4}
\usepackage{amsmath}
\usepackage{graphicx}
\usepackage{bm}
\usepackage{epsfig}

\makeatletter

\begin{document}

\title{Hund-Heisenberg model in superconducting infinite-layer nickelates}
\author{Jun Chang$^{1}$, Jize Zhao$^{2}$, Yang Ding$^{3}$}
\affiliation{$^{1}$College of Physics and Information Technology, Shaanxi Normal
University, Xi'an 710119, China~\\
 $^{2}$School of Physical Science and Technology\&Key Laboratory
for Magnetism and Magnetic Materials of the MoE, Lanzhou University,
Lanzhou 730000, China~\\
 $^{3}$Center for High-Pressure Science and Technology Advanced Research,
Beijing 100094, China~}
\date{\today} 

\begin{abstract}

We theoretically investigate the unconventional superconductivity
in the newly discovered infinite-layer nickelates Nd$_{1-x}$Sr$_{x}$NiO$_{2}$
based on a two-band model. By analyzing the transport experiments,
we propose that the doped holes dominantly enter the Ni $d_{xy}$
or/and $d_{3z^{2}-r^{2}}$ orbitals as charged carriers, and form a conducting band. Via
the onsite Hund coupling, the doped holes are coupled
to the Ni localized holes in the $d_{x^{2}-y^{2}}$ orbital band.
We demonstrate that this two-band model could be further reduced to
a Hund-Heisenberg model. Using the reduced model, we show the non-Fermi
liquid state above the critical $T_{c}$ could stem from the carriers
coupled to the spin fluctuations of the localized holes. In the superconducting
phase, the short-range spin fluctuations mediate the carriers into
Cooper pairs and establish $d_{x^{2}-y^{2}}$-wave superconductivity.
We further predict that the doped holes ferromagnetically coupled
with the local magnetic moments remain itinerant even at very low
temperature, and thus the pseudogap hardly emerges in nickelates.
Our work provides a new superconductivity mechanism for strongly correlated
multi-orbital systems and paves a distinct way to exploring new superconductors
in transition or rare-earth metal oxides. 
\end{abstract}

\maketitle

\section{Introduction}
Various unconventional superconductors have proliferated experimentally over the recent decades although the origins of the superconductivity (SC) in cuprates and heavy fermion materials remain theoretically controversial \cite{Bednorz1986,Steglich1979,Lee2006}. The quasi-two-dimensional
(2D) iron pnictides have triggered a new boom of the SC investigation
\cite{Hosono2008}. In particular, the search efforts for the compounds
with the geometrical and electronic structure similar to cuprates
are growing both experimentally and theoretically due to the highest
critical temperatures at ambient conditions. Isostructural compounds,
e.g. Sr$_{2}$RuO$_{4}$, Sr$_{2}$IrO$_{4}$ and LaNiO$_{2}$ as
well as some artificial heterostructure have been extensively proposed,
and synthetized \cite{maeno1994,li2019,Anisimov1999,Lee2004,yan2015,kim2016,chaloupka2008,schwingenschlogl2009stripe,hansmann2009,ikeda2016,kawai2009,kaneko2009,Ryee2019,Hirsch2019,Botana2019,hayward1999}.
The unremitting pursuits have been rewarded despite no evidence of
SC in some of these analogs so far.

Recently, the exciting discovery of superconductivity in the hole
doped infinite-layer nickelate Nd$_{1-x}$Sr$_{x}$NiO$_{2}$ redraws
strong attention to the unconventional SC \cite{li2019,Hepting2019,Sakakibara2019,Gao2019,bernardini2020magnetic,Jiang2019electronic}.
The quasi-2D Ni-O plane is geometrically analog to the Cu-O plane
in cuprates. The $d_{x^{2}-y^{2}}$ orbital of each Ni$^{1+}$ ion
is also half-filled, with an effective spin-1/2 on each site. However,
the differences from cuprates are notably striking. In the parent
compounds, there is no sign of long-range magnetic orders in the measured
temperature range \cite{hayward2003}. Maybe due to self-doping effects,
the electrons of the rare-earth Nd between Ni-O planes form a 3D weakly-interacting
$5d$ metallic state with an electronic Fermi surface \cite{HZhang2019,Wu2019,Hepting2019,Normura2019}.
Intriguingly, the resistivity exhibits metallic temperature dependence
down to 60 K, and then shows insulating upturn at lower temperatures,
which could be the results of weak localization effects, Kondo effects
or temperature driven intra-band transitions \cite{Singh2019,li2019,choi2020role}.
Upon chemical doping, additional holes dominantly enter the $d$ orbitals
of the Ni ions rather than O orbitals as in cuprates since the O $2p$
states are far away from the Fermi level in nickelates \cite{Lee2004,Jiang2019,HZhang2019,YHZhang2019,Gao2019}.
The sign change of the Hall coefficient at low temperature indicates
that both electrons and holes may contribute to the transport and
thermodynamic properties \cite{li2019}. Moreover, it is debating
whether the doped hole forms a spin singlet or triplet doublon with
the original hole on a Ni ion \cite{Jiang2019,Hu2019,YHZhang2019,Werner2019,GMZhang2019}.
Several microscopic models have been proposed, such as the $t$-$J$
models, the metallic gas coupled to a 2D Hubbard model and the spin
freezing model \cite{GMZhang2019,HZhang2019,fu2019corelevel,Wu2019,Hu2019,YHZhang2019,Hepting2019,Werner2019}.
More surprisingly, absence of superconductivity was recently claimed
in the bulk nickelates and the film prepared on various oxide substrates
different from SrTiO. It was suggested that the absence possibly
results from the hydrogen intercalation \cite{qLi2019,xZhou2019,si2019topotactic}.
Own to these confusions, more insights into the microscopic mechanism
in nickelates are imperative.

In this paper, we investigate the nickelate SC based on the analysis
of the transport experiments. Considering the positive Hall coefficient
and the suppressed self-doping effects at low temperature, we suggest
that  since the Ni $d_{xy}$ or/and $d_{3z^{2}-r^{2}}$
orbital is close to Fermi energy level, the doped holes may go to these orbitals and establish a conducting
band \cite{Lee2004,lechermann2019late,Sakakibara2019,Gao2019}. The
onsite Hund interaction couples the conducting band with the localized
$d_{x^{2}-y^{2}}$ orbital band together. The correlation between
the sparse carriers and the kinetic energy of the localized holes
could then be ignored. Thus, the two-band model is simplified into
a Hund-Heisenberg model. We show that both the non-Fermi liquid in normal
state and the superconductivity is determined by the spin fluctuations
of the localized holes. This SC mechanism could be realized in multi-orbital
strongly correlated systems with both Hund and Heisenberg interactions \cite{Lee2018,Georges2013,Haule2009,Werner2008}.

\section{Microscopic Hamiltonian}
We first analyze the electronic properties in normal state based on
the transport experiments \cite{li2019}. In the parent compounds,
both the resistivity and Hall effect measurements show Kondo effects with logarithmic temperature dependence from tens to around several
kelvin. The Kondo effects were attributed to the hybridization between
the Nd $5d$ states and the Nd $4f$ or Ni $3d$ states as in rare-earth
heavy fermion compounds although the $\textit{ab initio }$ study
suggests the hybridization between the Ni $3d$ state and the Nd $5d$
states is negligible, and the $4f$ electron spin fluctuation should
be weak due to the large magnetic moment and the energy far away from
Fermi energy level \cite{GMZhang2019,HZhang2019,Normura2019,Wu2019}.
In Nd$_{0.8}$Sr$_{0.2}$NiO$_{2}$, above 60 K, the negative Hall
coefficient indicates that the Nd $5d$ electrons dominate the transport
and thermodynamic properties. With the decreasing of temperature,
the self-doping effect is reduced as in semiconductors and the Hall
coefficient also changes its sign from negative to positive. This
means that the doped holes take over the dominant role in the transport
and thermodynamics at low temperature. In addition, the Hund coupling
between the $3d$ doped holes and localized holes is around an order
of magnitude stronger than the Kondo coupling between the $5d$ electron
and the $3d$ magnetic moments. Therefore, we ignore the $5d$ electrons
in our model, and in our discussion section we show that they only
give negligible contribution to the superconductivity and non-Fermi-liquid
behavior in the normal state. Whether the doped hole forms a high
spin triplet or a low spin singlet doublon with the original hole
on the $d_{x^{2}-y^{2}}$ orbital is still controversial \cite{Jiang2019,Hu2019,YHZhang2019,Werner2019,GMZhang2019}.
In fact, a Ni$^{2+}$ ion with $d^{8}$ configuration often has a
high spin S = 1 in common Nickel oxides as the result of Hund coupling.
According to the first principle calculation, the energy of the triplet
state is around 1 eV lower than that of the singlet, and the top of
the Ni $d_{xy}$ or/and $d_{3z^{2}-r^{2}}$ orbital band is also close
to the Fermi energy \cite{Lee2004,lechermann2019late,Sakakibara2019,Gao2019}.
Moreover, the holes doped on the $d_{xy}$ or/and $d_{3z^{2}-r^{2}}$
orbitals could itinerate freely, agreeing well with the positive Hall
coefficient at low temperature. The delocalization of the doped holes
on $d_{xy}$ or/and $d_{3z^{2}-r^{2}}$ orbitals is attributed to
the fact that the doped hole concentration is dilute, and under the
short range antiferromagetic (AF) correlation background, a hole can
hop freely to its next nearest neighbor sites without energy cost
as long as these sites are not occupied by another doped hole. In
contrast, the doped holes on the already half-filled $d_{x^{2}-y^{2}}$
orbitals tend to be localized at low temperature, otherwise the hopping
disturbs the magnetic configurations of short-range AF correlations
as in cuprates \cite{Lee2006}.

\begin{figure}[t]
\includegraphics[width=1\columnwidth]{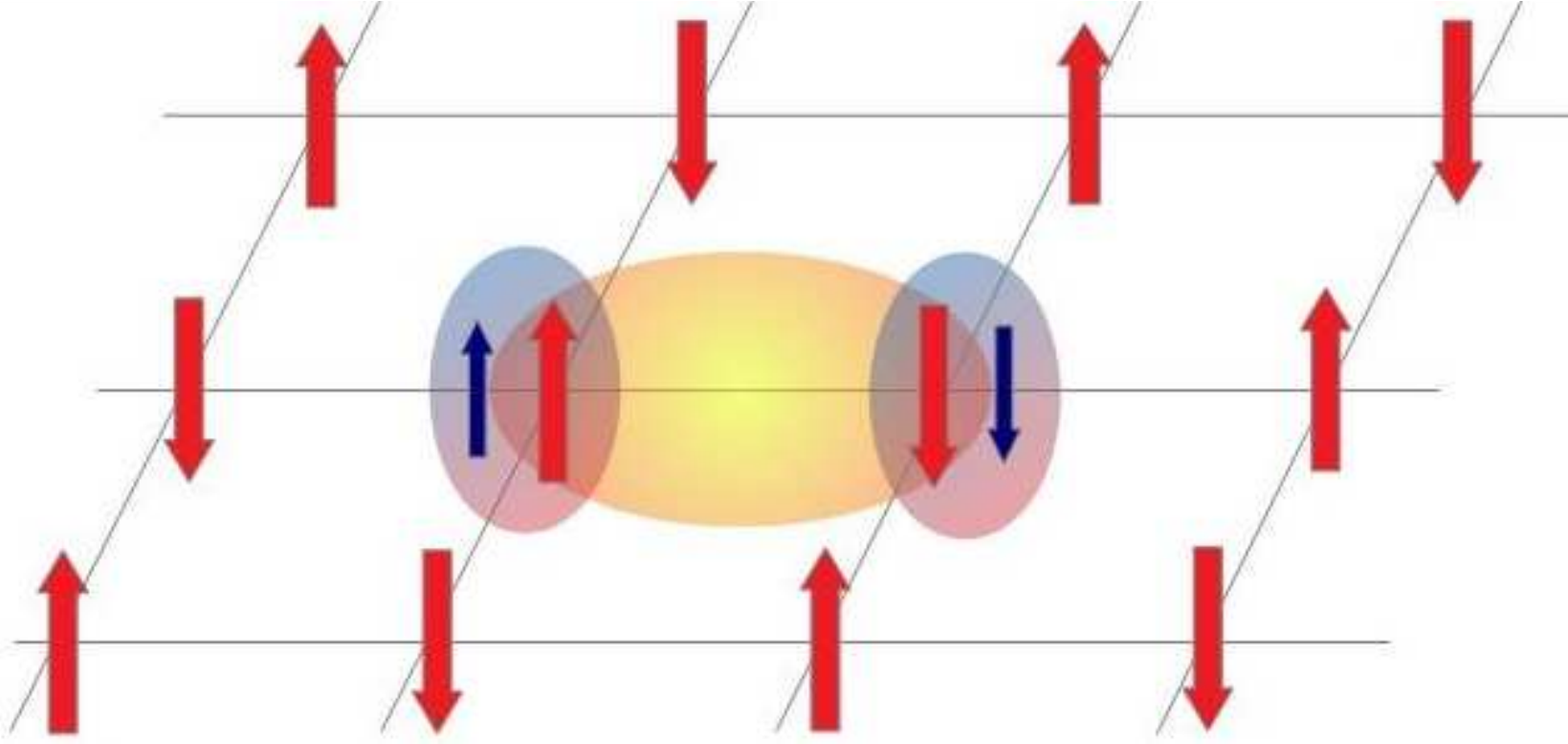} \caption{ Schematic spin configuration on the Ni-O planes of the hole-doped
nickenates. The red thicker arrows denote the spins of the localized holes
in the $d_{x^{2}-y^{2}}$ orbital band. The local magnetic moments
interact with each other through the Heisenberg interaction. The blue
arrows denotes the spins of the doped holes as carriers in the conducting
$d_{xy}$ or/and $d_{3z^{2}-r^{2}}$ orbital band. Their spins are
parallel with the local magnetic moments due to the strong Hund coupling.
In normal state, the scattering by the spin fluctuations of the localized
holes transfers the carrier Fermi gas into a non-Fermi liquid. A carrier
could hop to its next nearest neighbor sites without any energy cost
as long as these sites are not occupied by other doped holes. Thus
the doped holes itinerate over the lattice even at very low temperature
without formation of a pseudogap. While in superconducting phase,
two neighboring carrier particles are mediated into a Cooper pair
by the spin fluctuations of the local magnetic moments. \label{lattice} }
\end{figure}

Based on the aforementioned analysis, we confine our study to the
Ni-O planes and assume that the holes doped on the $d_{xy}$ or/and
$d_{3z^{2}-r^{2}}$ orbitals form a conducting band, coexisting with
the localized $d_{x^{2}-y^{2}}$ orbital band. $c_{i}=(c_{i\uparrow},c_{i\downarrow})^{T}$
is introduced as the annihilation operator of the bare carrier particles
on the $d_{xy}$ or/and $d_{3z^{2}-r^{2}}$ orbitals of the $i$th
Ni site, and $d_{i}=(d_{i\uparrow},d_{i\downarrow})^{T}$ is the real
space annihilation operator of the localized holes on the Ni $d_{x^{2}-y^{2}}$
orbitals. The Hamiltonian is written as 
\begin{eqnarray}
H & = & H_{c}+H_{d}+U_{c}\sum_{i}n_{ci\uparrow}n_{ci\downarrow}+U_{d}\sum_{i}n_{di\uparrow}n_{di\downarrow}\nonumber \\
 &  & +U_{cd}\sum_{i}n_{ci}n_{di}-J_{h}\sum_{i}\mathbf{S}_{ci}\cdot\mathbf{S}_{di},
\end{eqnarray}
with 
\begin{eqnarray}
H_{c}=\varepsilon_{c0}\sum_{i}c_{i}^{\dagger}c_{i}-\sum_{i,j}t_{cij}c_{i}^{\dagger}c_{j}\label{H0-2}
\end{eqnarray}
and 
\begin{eqnarray}
H_{d}=\varepsilon_{d0}\sum_{i}d_{i}^{\dagger}d_{i}-\sum_{i,j}t_{dij}d_{i}^{\dagger}d_{j},\label{H0-2-1}
\end{eqnarray}
where $t_{cij}$ and $t_{dij}$ are the hopping integrals of the carriers
in conducting band and the localized holes in $d_{x^{2}-y^{2}}$ band,
respectively. $n_{ci}=c_{i}^{\dagger}c_{i}=n_{ci\uparrow}+n_{ci\downarrow}$
is the occupancy of the carriers with spin up and down on the $i$th
site, and $n_{di}$ is the occupancy of the $d_{x^{2}-y^{2}}$ orbital.
$J_{h}$ is the Hund coupling between the carriers and the localized
holes on the same site. $U_{c}$ is the onsite Coulomb repulsion between
carriers, $U_{d}$ is the interaction between the localized holes,
and $U_{cd}$ is the inter-orbital Coulomb repulsion between the carrier
and localized hole on the same site. 

To further simplify the Hamiltonian, based on the fact that the hopping
term $t_{dij}$ is much less than the large onsite Coulomb interaction
$U_{d}$, we take the $t_{dij}$ as the perturbation then the Hubbard
model for the localized $d_{x^{2}-y^{2}}$ orbital band is reduced
to a Heisenberg model. In addition, the large onsite Coulomb interaction
strongly suppresses the particle number fluctuation on the half-filled
$d_{x^{2}-y^{2}}$ orbitals, so that the Coulomb interaction acting
on the carriers by the localized holes could be renormalized into
the chemical potential within the singly occupied approximation $\left\langle n_{di}\right\rangle =1$.
Furthermore, in consideration of the delocalization and the sparse
concentration $\left\langle n_{ci}\right\rangle $ of the doped holes
in conducting band, we can safely ignore the correlation between the
carriers as in common metals. Alternatively, the magnetic coupling
$\mathbf{S}_{ci}\cdot\mathbf{S}_{cj}$ term is ignored due to the much
weaker magnetization of the carriers than that of the localized holes.
Finally, we arrive at a Hund-Heisenberg model, 
\begin{eqnarray}
H=H_{c}-J_{h}\sum_{i}\mathbf{S}_{ci}\cdot\mathbf{S}_{di}+J_{H}\sum_{\left\langle i,j\right\rangle }\mathbf{S}_{di}\cdot\mathbf{S}_{dj}\label{H-1}
\end{eqnarray}
where $\mathbf{S}_{ci}=c_{i}^{\dagger}\boldsymbol{\sigma}c_{i}/2$
and $\mathbf{S}_{di}=d_{i}^{\dagger}\boldsymbol{\sigma}d_{i}/2$ with
the Pauli vector $\bm{\sigma}$. $J_{H}$ is the Heisenberg interaction
between the $d_{x^{2}-y^{2}}$ orbital holes on the Ni square lattice.
Here, we have ignored the kinetic energy of the localized holes, and
the carrier chemical potential $\varepsilon_{c0}$ has been replaced
by $\varepsilon_{c}=\varepsilon_{c0}+U_{cd}$ in $H_{c}$ to include
the energy renormalization from the Coulomb interaction of the holes
on the $d_{x^{2}-y^{2}}$ orbitals.

This model is formally similar to the Kondo-Heisenberg model on a
2D square lattice \cite{Chang2017,Zaanen1988}. The difference is
the conducting carrier ferromagnetically coupled to the localized
magnetic moments rather than antiferromangetically. In addition, the
onsite Hund ferromagnetic coupling is in favor of the delocalization
of the doped holes under the antiferromagnetic background even at
very low temperature, and thus it is difficult to form a pseudogap.
In the following, we show that the spin fluctuations of the localized
$d_{x^{2}-y^{2}}$ states not only act as the pairing `glue' in superconducting
state, but also results in non-Fermi liquid in normal state.

\section{Normal state}
The bare doped holes are assumed to compose a dilute Fermi gas with
the retarded Green's function $G_{c}^{0}(\mathbf{k},\omega)$, bathed
in the Heisenberg antiferromagnets of the localized holes. At low
temperature, the transport and thermodynamic properties are determined
by the imaginary part of the carrier self-energy. However, different
from the Fermi liquids, the dominant contribution to the self-energy is from the interaction between the carriers and the
localized holes rather than the carrier-carrier
interaction. Because the concentration of the carriers is much lower than the localized holes despite the same order strength of the two kinds of interactions. Therefore, we ignore
the self-energy correction by the carrier-carrier correlation interaction.
Within the Born approximation, the imaginary part of the self-energy
correction by the renormalized spin fluctuation $\chi_{d}(\mathbf{q},\omega)$
of the localized $d_{x^{2}-y^{2}}$ holes reads \cite{Chang2017}

\begin{eqnarray*}
 &  & \mbox{Im}\Sigma_{c}(\mathbf{k},\omega)\sim J_{h}^{2}\int_{-\omega_{c}}^{\omega_{c}}dv\left[n_{B}(v)+n_{F}(\omega+v)\right]I(\mathbf{k},\omega,v),
\end{eqnarray*}
where $n_{B}$ and $n_{F}$ are the Bose and Fermi functions, $\omega_{c}$
is the upper cutoff frequency of magnetic fluctuations and 
\begin{eqnarray}
I(\mathbf{k},\omega,v)=\int\frac{d^{2}q}{4\pi^{2}}\mbox{Im}\chi_{d}(\mathbf{q},v)\mbox{Im}G_{c}^{0}(\mathbf{k+q},\omega+v).\label{Ikwv}
\end{eqnarray}
It is worth noting that the neglect of the higher order self-energy
corrections is based on the fact that the carriers is weakly magnetized
by the local magnetic moments or $\left|\left\langle \mathbf{S}_{ci}\right\rangle \right|\ll1/2$.
Therefore, the Hund coupling in the effective Hamiltonian only gives
perturbation correction to the carrier self-energy despite the large
Hund coupling constant in Eq. (\ref{H-1}). However, the the approximation of neglecting the higher order corrections may be questionable if 
the $\mbox{Im}\chi_{d}(\mathbf{q},v)\sim v^s$ with $s\leq 0$ at low frequency. For instance, the perturbation related coupling constant $\lambda_\mathbf{q} =2\int_0^{\infty}dv\alpha_\mathbf{q}^2(v)\mbox{Im}\chi_{d}(\mathbf{q},v)/v$ diverges, where $\alpha_\mathbf{q}^2(v)\mbox{Im}\chi_{d}(\mathbf{q},v)$ is the generalized McMillan carrier-boson coupling function. 
 
Integrating over the momentum $\mathbf{k}$, one finds the momentum-integral
imaginary part of the Fermi gas self-energy
\begin{eqnarray*}
 &  & \mbox{Im}\Sigma_{c}(\omega)\sim J_{h}^{2}\int_{-\omega_{c}}^{\omega_{c}}dv\left[n_{B}(v)+n_{F}(\omega+v)\right]\mbox{Im}\chi_{d}(v)\rho_{c}(\omega+v)
\end{eqnarray*}
with the aid of 
\begin{eqnarray}
\int\frac{d^{2}k}{4\pi^{2}}I(\mathbf{k},\omega,v)=-\pi\rho_{c}(\omega+v)\mbox{Im}\chi_{d}(v)\label{I2}
\end{eqnarray}
where the magnetic fluctuations $\chi_{d}(v)\equiv\int d^{2}q\chi_{d}(\mathbf{q},v)/4\pi^{2}$.
$\rho_{c}$ is the density of states of the carriers. Since the absolute
value of $n_{B}(v)+n_{F}(\omega+v)$ exponentially decreases to zero
with increasing $|v|$, the energy range of integration can be extended
from the cutoff $\omega_{c}$ to infinity, and $\rho_{c}(\omega+v)$
is approximately substituted by $\rho_{c}^{0}$, the carrier density
of states at Fermi energy level. One has 
\begin{eqnarray*}
 &  & \mbox{Im}\Sigma_{c}(\omega)\sim J_{h}^{2}\rho_{c}^{0}\int_{-\infty}^{\infty}dv\left[n_{B}(v)+n_{F}(\omega+v)\right]\mbox{Im}\chi_{d}(v).
\end{eqnarray*}

Since no experimental or theoretical results on $\mbox{Im}\chi_{d}(v)$
could be obtained presently we assume that the momentum-integral spin-fluctuation
spectra of the half-filled $d_{x^{2}-y^{2}}$ holes take the similar
form of the underdoped cuprates\cite{Hayden1991,Keimer1991,Tranquada1992}
, e.g. $\mbox{Im}\chi_{d}(v)\sim\mbox{tanh}\left(v/2T\right)$. Using
the analytical frequency integral equation\cite{Chang2017} 

\begin{eqnarray}
\int_{-\infty}^{\infty}dv\left[n_{B}(v)+n_{F}(\omega+v)\right]\mbox{tanh}\left(\frac{v}{2T}\right)\nonumber \\
=2T\left[1+\frac{\omega}{2T}\mbox{tanh}\left(\frac{\omega}{2T}\right)\right],
\end{eqnarray}
the carriers have the marginal Fermi liquid-like self-energy 
\begin{eqnarray}
\mbox{Im}\Sigma_{c}(\omega,T) & \sim & \pi\rho_{c}^{0}J_{h}^{2}T\left[1+\frac{\omega}{2T}\mbox{tanh}\left(\frac{\omega}{2T}\right)\right]\nonumber \\
 & \sim & \max{(|\omega|,T)}.\label{selfenergy2}
\end{eqnarray}
Then the linear temperature dependence of electrical resistivity observed
in the experiments could be explained \cite{li2019}, and some other
anomalous transport properties are expected to be experimentally verified.

In cuprates, since the doped holes are antiferromagnetically coupled
to the localized holes, the hopping disturbs the original magnetic
configuration, and thus the doped holes tend to be localized at low
doping or at low temperature \cite{Lee2006}. On the contrary, in
nickelates, the doped holes, which are ferromagnetically coupled to
localized holes, could hop over the Ni-O plane without affecting the
magnetic background so that they are not easy to be trapped around
the local magnetic moments even at low temperature. In addition, a carrier could hop to its next nearest neighbor sites without any energy cost as long as these sites are not occupied by other doped holes. Therefore, we
propose that it is almost impossible to observe a pseudogap in nickelates.

\section{Superconductivity}

In normal state, the carriers are scattered by the short-range spin
fluctuation as the metallic gas by phonons. In the superconducting
state, the carrier pairing is mediated by the spin fluctuations analogues
of the pairing by phonons in the BCS mechanism. It is worthy of note
that we assume that the conducting carriers only partially screen
the local moments without formation of localized triplets or singlets,
and then the Heisenberg interaction between the screened moments and
their surroundings could survive. Thus, the carriers on unit cell
$i$ and $j$ could interact with each other by exchange of the spin
fluctuation in terms of a four point vertex, written in real space
as 
\begin{eqnarray}
\Gamma_{\alpha\beta,\gamma\delta}(i,j,\omega)=-\frac{J_{h}^{2}}{4}\chi_{d}(i,j,\omega)\sigma_{\alpha\beta}\sigma_{\gamma\delta}.\label{4vertex}
\end{eqnarray}
We assume that the AF spin fluctuation only mediates the itinerant holes spacing within the AF correlation length into stable Cooper pairs. The correlation length in nickelates is assumed to be the same scale as that in cuprates at low temperature, around two times the lattice constant.
Thus, we only take the nearest-neighbor $\chi_{d}(\left\langle i,j\right\rangle ,\omega)$
into account. Then, the interaction Hamiltonian of the carriers can
be written in the coordinate representation as \cite{Chang2017} 
\begin{eqnarray}
H_{sc}=J_{h}^{2}\chi_{d}(\left\langle i,j\right\rangle ,\omega)\sum_{\left\langle i,j\right\rangle }\mathbf{S}_{ci}\cdot\mathbf{S}_{cj},\label{Hsc-1}
\end{eqnarray}
where the nearest neighbor $\chi_{d}(\left\langle i,j\right\rangle ,\omega)$
is assumed to be space independent. For a local pair, the energy of
a spin-triplet is about $J_{h}^{2}\chi_{d}(\left\langle i,j\right\rangle ,\omega)$
higher than that of a spin-singlet, and thus the antiferromagnetic
spin fluctuations favors spin-singlet pairing. Combining with $H_{c}$
in Eq.~(\ref{H0-2}), a $t$-$J$-like model is reached. Interestingly,
despite the formal similarity with the conventional $t$-$J$ model
\cite{Zhang1988}, here the spin-like operator $\mathbf{s}$ is associated
with the carriers rather than the localized holes. 

After transforming to momentum space, the Hamiltonian on the square
lattice becomes 
\begin{eqnarray}
H_{sc}=\int\frac{d^{2}kd^{2}k'}{(2\pi)^{4}}J\left({\rm \mathbf{k}}-{\rm \mathbf{k}}'\right)c_{{\rm \mathbf{k}}\uparrow}^{\dagger}c_{-{\rm \mathbf{k}}\downarrow}^{\dagger}c_{-{\rm \mathbf{k}}'\downarrow}c_{{\rm \mathbf{k}}'\uparrow},\label{hint}
\end{eqnarray}
with 
\begin{eqnarray}
J\left({\rm \mathbf{k}}-{\rm \mathbf{k}}'\right)=-2g\left[\cos\left(k_{x}-k'_{x}\right)+\cos\left(k_{y}-k'_{y}\right)\right]
\end{eqnarray}
and the effective coupling between the carriers 
\begin{eqnarray}
g\equiv\frac{3}{4}J_{h}^{2}\chi_{d}(\left\langle i,j\right\rangle ,\omega).
\end{eqnarray}
The Cooper pairing potentials are symmetrized with $J({\rm \mathbf{k}}-{\rm \mathbf{k}}')$
and $J({\rm \mathbf{k}}+{\rm \mathbf{k}}')$ in the singlet channel
as \cite{Coleman2007}

\begin{eqnarray}
V_{{\rm \mathbf{k}},{\rm \mathbf{k}}'}=\frac{J\left({\rm \mathbf{k}}-{\rm \mathbf{k}}'\right)+J\left({\rm \mathbf{k}}+{\rm \mathbf{k}}'\right)}{2},
\end{eqnarray}
i.e. 
\begin{eqnarray}
V_{{\rm \mathbf{k}},{\rm \mathbf{k}}'}= & -2g\left[\cos k_{x}\cos k'_{x}+\cos k_{y}\cos k'_{y}\right]
\end{eqnarray}

The pairing interaction can be further decoupled into $d$-wave and
$s$-wave components, 
\begin{eqnarray}
2\cos k_{x}\cos k'_{x}+2\cos k_{y}\cos k'_{y}=\gamma_{k}\gamma_{k'}+\gamma_{k}^{s}\gamma_{k'}^{s},
\end{eqnarray}
with the $d$-wave gap function$\gamma_{k}=\cos k_{x}-\cos k_{y}$,
and the extended $s$-wave gap function $\gamma_{k}^{s}=\cos k_{x}+\cos k_{y}$.
For $s$-wave superconductivity, the pairing interaction is expected
to be negative and nearly isotropic. However, the interaction $V_{{\rm \mathbf{k}},{\rm \mathbf{k}}'}$
is positive at $\mathbf{k-k'\sim Q}$, and the AF spin fluctuation
is strongly momentum dependent, i.e. peaked at or near the AF wave vector $\mathbf{Q}$ \cite{Wu2019}. Therefore, only the $d$-wave pairing
channel is favored in the spin- fluctuation SC mechanism \cite{Scalapino1986,Bickers1987,Inui1988,Dong1988,Kotliar1988,Monthoux1991,Moriya1990,Millis1990,Chubukov2008}
and the attractive pairing interaction dominantly mediates the carriers
on the nearest-neighbor unit cells \cite{Scalapino2012}. Thus, the
pairing interaction is $V_{{\rm \mathbf{k}},{\rm \mathbf{k}}'}^{d}=-g\gamma_{k}\gamma_{k'}$
in the $d_{x^{2}-y^{2}}$ channel, which could be detected in phase sensitive interference measurements \cite{bker2020phasesensitive}. Consequently, in momentum space,
a weak coupling BCS interaction can be written 
\begin{eqnarray}
H_{scd}=-g\int\frac{d^{2}k}{4\pi^{2}}\gamma_{\mathbf{k}}c_{\mathbf{k}\uparrow}^{\dagger}c_{-\mathbf{k}\downarrow}^{\dagger}\int\frac{d^{2}k'}{4\pi^{2}}\gamma_{\mathbf{k'}}c_{\mathbf{k'}\uparrow}c_{-\mathbf{k'}\downarrow},
\end{eqnarray}
The superconductivity order parameter $\gamma_{\mathbf{k}}\Delta_{sc}$
is introduced in the mean field method with the BCS gap equation 
\begin{eqnarray}
\Delta_{sc}=-g\int\frac{d^{2}k}{4\pi^{2}}\gamma_{\mathbf{k}}<c_{\mathbf{k}\uparrow}^{\dagger}c_{-\mathbf{k}\downarrow}^{\dagger}>,
\end{eqnarray}
where $c_{\mathbf{k}\uparrow}^{\dagger}c_{-\mathbf{k}\downarrow}^{\dagger}$
is the Cooper pair operator denoting a bond state of two carriers
with opposite momentum and spin. 

Solving the gap equation in the limit $\Delta_{sc}(T\rightarrow T_{c})\rightarrow0$,
the SC transition temperature $T_{c}\sim\omega_{c}e^{-1/\lambda}$
with $\lambda=g\rho_{c}^{0}/2$ for weak coupling $d$-wave superconductors.
Since the AF correlation in nickelates is weaker than that in cuprates,
the spin fluctuation cutoff energy $\hbar\omega_{c}\sim J_H$ should be lower
than that in cuprates. Moreover, the Hund coupling $J_{h}$ is smaller
than the magnetic coupling $J_{K}$ between the O carriers and Cu
local moments in cuprates. Therefore, the lower critical temperature
$T_{c}$ in nickelates could be understood.

\section{Discussion and Conclusion}
Actually, we could not exclude the possibility that the doped holes
go to the Ni $d_{x^{2}-y^{2}}$ orbitals although the strong onsite
Coulomb repulsion pushes the $d_{x^{2}-y^{2}}$ lower Hubbard band
away from the Fermi level \cite{Lee2004,gu2019hybridization}. Nevertheless,
if the doped holes forms onsite spin singlet on the $d_{x^{2}-y^{2}}$
orbital with the original localized hole, the already weak AF coupling
is further suppressed and hence the superconductivity is weakened. The
critical temperature should also be sensitive to the doping level.
In addition, pseudogap should emerge at low temperature as in cuprates.
On the contrary, the doped holes on the $d_{xy}$ or/and $d_{3z^{2}-r^{2}}$
orbital may enhance the original AF couplings although the average
spin of a carrier is very weak. In addition, the SC critical temperature
depending on the Hund coupling $J_{h}$ and the spin fluctuation $\chi_{d}$
is not directly related to the doping level unless the carrier
density is too low. It is not likely to form a pseudogap due to
the ferromagnetic coupling between the carriers and the local magnetic
moments in our model. We expect more experiments to check these differences.

The Nd $5d$ electrons have been ignored in our model for the doped
holes dominate the transport and thermodynamics at low temperature
in the high doped examples. Nevertheless, in the recent paper\cite{li2020superconducting},
it was found that the Hall coefficients become negative at the doping
$x$ below 0.175, which means that electrons may be the dominant carriers
at low doping. The $5d$ electron as carrier could couple to the Ni
localized $d_{x^{2}-y^{2}}$ magnetic moments via the Kondo interaction\cite{GMZhang2019}.
Thus, the interaction between the $5d$ electrons and spin fluctuation
in nickelates could be described by the Kondo-Heisenberg model\cite{Chang2017}.
However, since the Kondo coupling ($J_{K}\sim$0.1 eV) is around ten
times smaller than the Hund coupling ($J_{h}\sim$ 1 eV), the $5d$
electrons give much weaker contribution to the superconductivity,
and their selfenergy renormalization in the normal state is also much
weaker than that of the $3d$ holes, namely, $T_{c}\sim\omega_{c}e^{-1/\lambda}$
with $\lambda\sim J_{K}^{2}$, $J_{h}^{2}$ and $\mbox{Im}\Sigma(\omega,T)\sim J_{K}^{2}$, $J_{h}^{2}$, respectively.

We have assumed that the momentum-integral spin-fluctuation spectra
of localized Ni localized $d_{x^{2}-y^{2}}$ holes on the Ni-O planes
take the similar form of the underdoped cuprates, and then the marginal
Fermi liquid-like self-energy is obtained. We expect that the neutron
scattering and more transport experiments could be conducted to verify
our assumption. Moreover, if the doped holes enter the $d_{xy}$ or/and
$d_{3z^{2}-r^{2}}$ orbital then the doping does not suppress the
AF fluctuations. This also could be judged by the neutron scattering
measurements. In addition, to experimentally determine if the doped
holes enter the $d_{xy}$ or/and $d_{3z^{2}-r^{2}}$ orbital, one
way is to apply Ni L-edge polarized x-ray absorption near edge structure
(XANES) on the single-crystals to study the distribution of holes
in the Ni $3d$ orbitals \cite{kaindl1989correlation}.

In conclusion, we have proposed a Hund-Heisenberg model to investigate
the unconventional SC in the infinite-layer nickelates superconductor.
By analyzing the transport experiments, we suggest that the doped
holes enter the Ni $d_{xy}$ or/and $d_{3z^{2}-r^{2}}$ orbitals,
and form a conducting band. The doped holes interact with the localized
holes on $d_{x^{2}-y^{2}}$ orbital through the onsite Hund coupling.
We show that the non-Fermi liquid state in normal phase results from
the carrier gas interacting with the spin fluctuations of the localized
holes. In the superconducting phase, it is still the short-range spin
fluctuations that mediate the carriers into Cooper pairs and leads
to $d$-wave superconductivity. We expect experiments to check our
predictions that the doped holes slightly enhance the spin fluctuations
and a pseudogap hardly forms in nickelates. We have provided a new
SC mechanism for multi-orbital strongly correlated systems, e.g. iron pnictides and it
should aid in probing or synthesizing new superconductors in transition
or rare-earth metal oxides.

\textit{Acknowledgements}$-$ We are thankful to Xun-Wang Yan, Yuehua Su and Myung Joon Han for
fruitful discussions. This work is supported by the National Natural Science Foundation
of China (91750111, 11874188, U1930401 and 11874075), and  National Key Research and Development Program of China (2018YFA0305703) and Science Challenge Project (TZ2016001).


\begin{thebibliography}{66}
\expandafter\ifx\csname natexlab\endcsname\relax\def\natexlab#1{#1}\fi
\expandafter\ifx\csname bibnamefont\endcsname\relax
  \def\bibnamefont#1{#1}\fi
\expandafter\ifx\csname bibfnamefont\endcsname\relax
  \def\bibfnamefont#1{#1}\fi
\expandafter\ifx\csname citenamefont\endcsname\relax
  \def\citenamefont#1{#1}\fi
\expandafter\ifx\csname url\endcsname\relax
  \def\url#1{\texttt{#1}}\fi
\expandafter\ifx\csname urlprefix\endcsname\relax\def\urlprefix{URL }\fi
\providecommand{\bibinfo}[2]{#2}
\providecommand{\eprint}[2][]{\url{#2}}

\bibitem[{\citenamefont{Bednorz and M\"{u}ller}(1986)}]{Bednorz1986}
\bibinfo{author}{\bibfnamefont{J.~G.} \bibnamefont{Bednorz}} \bibnamefont{and}
  \bibinfo{author}{\bibfnamefont{K.~A.} \bibnamefont{M\"{u}ller}},
  \bibinfo{journal}{Z. Phys. B} \textbf{\bibinfo{volume}{64}},
  \bibinfo{pages}{189} (\bibinfo{year}{1986}).

\bibitem[{\citenamefont{Steglich et~al.}(1979)\citenamefont{Steglich, Aarts,
  Bredl, Lieke, Meschede, Franz, and Sch\"afer}}]{Steglich1979}
\bibinfo{author}{\bibfnamefont{F.}~\bibnamefont{Steglich}},
  \bibinfo{author}{\bibfnamefont{J.}~\bibnamefont{Aarts}},
  \bibinfo{author}{\bibfnamefont{C.~D.} \bibnamefont{Bredl}},
  \bibinfo{author}{\bibfnamefont{W.}~\bibnamefont{Lieke}},
  \bibinfo{author}{\bibfnamefont{D.}~\bibnamefont{Meschede}},
  \bibinfo{author}{\bibfnamefont{W.}~\bibnamefont{Franz}}, \bibnamefont{and}
  \bibinfo{author}{\bibfnamefont{H.}~\bibnamefont{Sch\"afer}},
  \bibinfo{journal}{Phys. Rev. Lett.} \textbf{\bibinfo{volume}{43}},
  \bibinfo{pages}{1892} (\bibinfo{year}{1979}).

\bibitem[{\citenamefont{Lee et~al.}(2006)\citenamefont{Lee, Nagaosa, and
  Wen}}]{Lee2006}
\bibinfo{author}{\bibfnamefont{P.~A.} \bibnamefont{Lee}},
  \bibinfo{author}{\bibfnamefont{N.}~\bibnamefont{Nagaosa}}, \bibnamefont{and}
  \bibinfo{author}{\bibfnamefont{X.-G.} \bibnamefont{Wen}},
  \bibinfo{journal}{Rev. Mod. Phys.} \textbf{\bibinfo{volume}{78}},
  \bibinfo{pages}{17} (\bibinfo{year}{2006}).

\bibitem[{\citenamefont{Kamihara et~al.}(2008)\citenamefont{Kamihara, Watanabe,
  Hirano, and Hosono}}]{Hosono2008}
\bibinfo{author}{\bibfnamefont{Y.}~\bibnamefont{Kamihara}},
  \bibinfo{author}{\bibfnamefont{T.}~\bibnamefont{Watanabe}},
  \bibinfo{author}{\bibfnamefont{M.}~\bibnamefont{Hirano}}, \bibnamefont{and}
  \bibinfo{author}{\bibfnamefont{H.}~\bibnamefont{Hosono}},
  \bibinfo{journal}{J. Am. Chem. Soc.} \textbf{\bibinfo{volume}{130}},
  \bibinfo{pages}{3296} (\bibinfo{year}{2008}).

\bibitem[{\citenamefont{Maeno et~al.}(1994)\citenamefont{Maeno, Hashimoto,
  Yoshida, Nishizaki, Fujita, Bednorz, and Lichtenberg}}]{maeno1994}
\bibinfo{author}{\bibfnamefont{Y.}~\bibnamefont{Maeno}},
  \bibinfo{author}{\bibfnamefont{H.}~\bibnamefont{Hashimoto}},
  \bibinfo{author}{\bibfnamefont{K.}~\bibnamefont{Yoshida}},
  \bibinfo{author}{\bibfnamefont{S.}~\bibnamefont{Nishizaki}},
  \bibinfo{author}{\bibfnamefont{T.}~\bibnamefont{Fujita}},
  \bibinfo{author}{\bibfnamefont{J.}~\bibnamefont{Bednorz}}, \bibnamefont{and}
  \bibinfo{author}{\bibfnamefont{F.}~\bibnamefont{Lichtenberg}},
  \bibinfo{journal}{Nature} \textbf{\bibinfo{volume}{372}},
  \bibinfo{pages}{532} (\bibinfo{year}{1994}).

\bibitem[{\citenamefont{Li et~al.}(2019)\citenamefont{Li, Lee, Wang, Osada,
  Crossley, Lee, Cui, Hikita, and Hwang}}]{li2019}
\bibinfo{author}{\bibfnamefont{D.}~\bibnamefont{Li}},
  \bibinfo{author}{\bibfnamefont{K.}~\bibnamefont{Lee}},
  \bibinfo{author}{\bibfnamefont{B.~Y.} \bibnamefont{Wang}},
  \bibinfo{author}{\bibfnamefont{M.}~\bibnamefont{Osada}},
  \bibinfo{author}{\bibfnamefont{S.}~\bibnamefont{Crossley}},
  \bibinfo{author}{\bibfnamefont{H.~R.} \bibnamefont{Lee}},
  \bibinfo{author}{\bibfnamefont{Y.}~\bibnamefont{Cui}},
  \bibinfo{author}{\bibfnamefont{Y.}~\bibnamefont{Hikita}}, \bibnamefont{and}
  \bibinfo{author}{\bibfnamefont{H.~Y.} \bibnamefont{Hwang}},
  \bibinfo{journal}{Nature} \textbf{\bibinfo{volume}{572}},
  \bibinfo{pages}{624} (\bibinfo{year}{2019}).

\bibitem[{\citenamefont{Anisimov et~al.}(1999)\citenamefont{Anisimov,
  Bukhvalov, and Rice}}]{Anisimov1999}
\bibinfo{author}{\bibfnamefont{V.~I.} \bibnamefont{Anisimov}},
  \bibinfo{author}{\bibfnamefont{D.}~\bibnamefont{Bukhvalov}},
  \bibnamefont{and} \bibinfo{author}{\bibfnamefont{T.~M.} \bibnamefont{Rice}},
  \bibinfo{journal}{Phys. Rev. B} \textbf{\bibinfo{volume}{59}},
  \bibinfo{pages}{7901} (\bibinfo{year}{1999}).

\bibitem[{\citenamefont{Lee and Pickett}(2004)}]{Lee2004}
\bibinfo{author}{\bibfnamefont{K.-W.} \bibnamefont{Lee}} \bibnamefont{and}
  \bibinfo{author}{\bibfnamefont{W.~E.} \bibnamefont{Pickett}},
  \bibinfo{journal}{Phys. Rev. B} \textbf{\bibinfo{volume}{70}},
  \bibinfo{pages}{165109} (\bibinfo{year}{2004}).

\bibitem[{\citenamefont{Yan et~al.}(2015)\citenamefont{Yan, Ren, Xu, Xie, Tao,
  Choi, Lee, Choi, Zhang, and Feng}}]{yan2015}
\bibinfo{author}{\bibfnamefont{Y.~J.} \bibnamefont{Yan}},
  \bibinfo{author}{\bibfnamefont{M.~Q.} \bibnamefont{Ren}},
  \bibinfo{author}{\bibfnamefont{H.~C.} \bibnamefont{Xu}},
  \bibinfo{author}{\bibfnamefont{B.~P.} \bibnamefont{Xie}},
  \bibinfo{author}{\bibfnamefont{R.}~\bibnamefont{Tao}},
  \bibinfo{author}{\bibfnamefont{H.~Y.} \bibnamefont{Choi}},
  \bibinfo{author}{\bibfnamefont{N.}~\bibnamefont{Lee}},
  \bibinfo{author}{\bibfnamefont{Y.~J.} \bibnamefont{Choi}},
  \bibinfo{author}{\bibfnamefont{T.}~\bibnamefont{Zhang}}, \bibnamefont{and}
  \bibinfo{author}{\bibfnamefont{D.~L.} \bibnamefont{Feng}},
  \bibinfo{journal}{Phys. Rev. X} \textbf{\bibinfo{volume}{5}},
  \bibinfo{pages}{041018} (\bibinfo{year}{2015}).

\bibitem[{\citenamefont{Kim et~al.}(2016)\citenamefont{Kim, Sung, Denlinger,
  and Kim}}]{kim2016}
\bibinfo{author}{\bibfnamefont{Y.~K.} \bibnamefont{Kim}},
  \bibinfo{author}{\bibfnamefont{N.}~\bibnamefont{Sung}},
  \bibinfo{author}{\bibfnamefont{J.}~\bibnamefont{Denlinger}},
  \bibnamefont{and} \bibinfo{author}{\bibfnamefont{B.}~\bibnamefont{Kim}},
  \bibinfo{journal}{Nature Phys.} \textbf{\bibinfo{volume}{12}},
  \bibinfo{pages}{37} (\bibinfo{year}{2016}).

\bibitem[{\citenamefont{Chaloupka and Khaliullin}(2008)}]{chaloupka2008}
\bibinfo{author}{\bibfnamefont{J.}~\bibnamefont{Chaloupka}} \bibnamefont{and}
  \bibinfo{author}{\bibfnamefont{G.}~\bibnamefont{Khaliullin}},
  \bibinfo{journal}{Phys. Rev. Lett.} \textbf{\bibinfo{volume}{100}},
  \bibinfo{pages}{016404} (\bibinfo{year}{2008}).

\bibitem[{\citenamefont{Schwingenschl{\"o}gl
  et~al.}(2009)\citenamefont{Schwingenschl{\"o}gl, Schuster, and
  Fr{\'e}sard}}]{schwingenschlogl2009stripe}
\bibinfo{author}{\bibfnamefont{U.}~\bibnamefont{Schwingenschl{\"o}gl}},
  \bibinfo{author}{\bibfnamefont{C.}~\bibnamefont{Schuster}}, \bibnamefont{and}
  \bibinfo{author}{\bibfnamefont{R.}~\bibnamefont{Fr{\'e}sard}},
  \bibinfo{journal}{Annalen der Physik} \textbf{\bibinfo{volume}{18}},
  \bibinfo{pages}{107} (\bibinfo{year}{2009}).

\bibitem[{\citenamefont{Hansmann et~al.}(2009)\citenamefont{Hansmann, Yang,
  Toschi, Khaliullin, Andersen, and Held}}]{hansmann2009}
\bibinfo{author}{\bibfnamefont{P.}~\bibnamefont{Hansmann}},
  \bibinfo{author}{\bibfnamefont{X.}~\bibnamefont{Yang}},
  \bibinfo{author}{\bibfnamefont{A.}~\bibnamefont{Toschi}},
  \bibinfo{author}{\bibfnamefont{G.}~\bibnamefont{Khaliullin}},
  \bibinfo{author}{\bibfnamefont{O.~K.} \bibnamefont{Andersen}},
  \bibnamefont{and} \bibinfo{author}{\bibfnamefont{K.}~\bibnamefont{Held}},
  \bibinfo{journal}{Phys. Rev. Lett.} \textbf{\bibinfo{volume}{103}},
  \bibinfo{pages}{016401} (\bibinfo{year}{2009}).

\bibitem[{\citenamefont{Ikeda et~al.}(2016)\citenamefont{Ikeda, Krockenberger,
  Irie, Naito, and Yamamoto}}]{ikeda2016}
\bibinfo{author}{\bibfnamefont{A.}~\bibnamefont{Ikeda}},
  \bibinfo{author}{\bibfnamefont{Y.}~\bibnamefont{Krockenberger}},
  \bibinfo{author}{\bibfnamefont{H.}~\bibnamefont{Irie}},
  \bibinfo{author}{\bibfnamefont{M.}~\bibnamefont{Naito}}, \bibnamefont{and}
  \bibinfo{author}{\bibfnamefont{H.}~\bibnamefont{Yamamoto}},
  \bibinfo{journal}{Appl. Phys. Express} \textbf{\bibinfo{volume}{9}},
  \bibinfo{pages}{061101} (\bibinfo{year}{2016}).

\bibitem[{\citenamefont{Kawai et~al.}(2009)\citenamefont{Kawai, Inoue,
  Mizumaki, Kawamura, Ichikawa, and Shimakawa}}]{kawai2009}
\bibinfo{author}{\bibfnamefont{M.}~\bibnamefont{Kawai}},
  \bibinfo{author}{\bibfnamefont{S.}~\bibnamefont{Inoue}},
  \bibinfo{author}{\bibfnamefont{M.}~\bibnamefont{Mizumaki}},
  \bibinfo{author}{\bibfnamefont{N.}~\bibnamefont{Kawamura}},
  \bibinfo{author}{\bibfnamefont{N.}~\bibnamefont{Ichikawa}}, \bibnamefont{and}
  \bibinfo{author}{\bibfnamefont{Y.}~\bibnamefont{Shimakawa}},
  \bibinfo{journal}{Appl. Phys. Lett.} \textbf{\bibinfo{volume}{94}},
  \bibinfo{pages}{082102} (\bibinfo{year}{2009}).

\bibitem[{\citenamefont{Kaneko et~al.}(2009)\citenamefont{Kaneko, Yamagishi,
  Tsukada, Manabe, and Naito}}]{kaneko2009}
\bibinfo{author}{\bibfnamefont{D.}~\bibnamefont{Kaneko}},
  \bibinfo{author}{\bibfnamefont{K.}~\bibnamefont{Yamagishi}},
  \bibinfo{author}{\bibfnamefont{A.}~\bibnamefont{Tsukada}},
  \bibinfo{author}{\bibfnamefont{T.}~\bibnamefont{Manabe}}, \bibnamefont{and}
  \bibinfo{author}{\bibfnamefont{M.}~\bibnamefont{Naito}},
  \bibinfo{journal}{Physica C: Superconductivity}
  \textbf{\bibinfo{volume}{469}}, \bibinfo{pages}{936} (\bibinfo{year}{2009}).

\bibitem[{\citenamefont{Ryee et~al.}(2020)\citenamefont{Ryee, Yoon, Kim, Jeong,
  and Han}}]{Ryee2019}
\bibinfo{author}{\bibfnamefont{S.}~\bibnamefont{Ryee}},
  \bibinfo{author}{\bibfnamefont{H.}~\bibnamefont{Yoon}},
  \bibinfo{author}{\bibfnamefont{T.~J.} \bibnamefont{Kim}},
  \bibinfo{author}{\bibfnamefont{M.~Y.} \bibnamefont{Jeong}}, \bibnamefont{and}
  \bibinfo{author}{\bibfnamefont{M.~J.} \bibnamefont{Han}},
  \bibinfo{journal}{Phys. Rev. B} \textbf{\bibinfo{volume}{101}},
  \bibinfo{pages}{064513} (\bibinfo{year}{2020}).

\bibitem[{\citenamefont{Hirsch and Marsiglio}(2019)}]{Hirsch2019}
\bibinfo{author}{\bibfnamefont{J.}~\bibnamefont{Hirsch}} \bibnamefont{and}
  \bibinfo{author}{\bibfnamefont{F.}~\bibnamefont{Marsiglio}},
  \bibinfo{journal}{Physica C (Amsterdam)} \textbf{\bibinfo{volume}{566}},
  \bibinfo{pages}{1353534} (\bibinfo{year}{2019}), ISSN
  \bibinfo{issn}{0921-4534}.

\bibitem[{\citenamefont{Botana and Norman}(2020)}]{Botana2019}
\bibinfo{author}{\bibfnamefont{A.~S.} \bibnamefont{Botana}} \bibnamefont{and}
  \bibinfo{author}{\bibfnamefont{M.~R.} \bibnamefont{Norman}},
  \bibinfo{journal}{Phys. Rev. X} \textbf{\bibinfo{volume}{10}},
  \bibinfo{pages}{11024} (\bibinfo{year}{2020}).

\bibitem[{\citenamefont{Hayward et~al.}(1999)\citenamefont{Hayward, Green,
  Rosseinsky, and Sloan}}]{hayward1999}
\bibinfo{author}{\bibfnamefont{M.}~\bibnamefont{Hayward}},
  \bibinfo{author}{\bibfnamefont{M.}~\bibnamefont{Green}},
  \bibinfo{author}{\bibfnamefont{M.}~\bibnamefont{Rosseinsky}},
  \bibnamefont{and} \bibinfo{author}{\bibfnamefont{J.}~\bibnamefont{Sloan}},
  \bibinfo{journal}{J. Am. Chem. Soc.} \textbf{\bibinfo{volume}{121}},
  \bibinfo{pages}{8843} (\bibinfo{year}{1999}).

\bibitem[{\citenamefont{{Hepting} et~al.}(2020)\citenamefont{{Hepting}, {Li},
  {Jia}, {Lu}, {Paris}, {Tseng}, {Feng}, {Osada}, {Been}, {Hikita}
  et~al.}}]{Hepting2019}
\bibinfo{author}{\bibfnamefont{M.}~\bibnamefont{{Hepting}}},
  \bibinfo{author}{\bibfnamefont{D.}~\bibnamefont{{Li}}},
  \bibinfo{author}{\bibfnamefont{C.~J.} \bibnamefont{{Jia}}},
  \bibinfo{author}{\bibfnamefont{H.}~\bibnamefont{{Lu}}},
  \bibinfo{author}{\bibfnamefont{E.}~\bibnamefont{{Paris}}},
  \bibinfo{author}{\bibfnamefont{Y.}~\bibnamefont{{Tseng}}},
  \bibinfo{author}{\bibfnamefont{X.}~\bibnamefont{{Feng}}},
  \bibinfo{author}{\bibfnamefont{M.}~\bibnamefont{{Osada}}},
  \bibinfo{author}{\bibfnamefont{E.}~\bibnamefont{{Been}}},
  \bibinfo{author}{\bibfnamefont{Y.}~\bibnamefont{{Hikita}}},
  \bibnamefont{et~al.}, \bibinfo{journal}{Nat. Mater.}
  \textbf{\bibinfo{volume}{19}}, \bibinfo{pages}{381} (\bibinfo{year}{2020}).

\bibitem[{\citenamefont{Sakakibara et~al.}(2019)\citenamefont{Sakakibara, Usui,
  Suzuki, Kotani, Aoki, and Kuroki}}]{Sakakibara2019}
\bibinfo{author}{\bibfnamefont{H.}~\bibnamefont{Sakakibara}},
  \bibinfo{author}{\bibfnamefont{H.}~\bibnamefont{Usui}},
  \bibinfo{author}{\bibfnamefont{K.}~\bibnamefont{Suzuki}},
  \bibinfo{author}{\bibfnamefont{T.}~\bibnamefont{Kotani}},
  \bibinfo{author}{\bibfnamefont{H.}~\bibnamefont{Aoki}}, \bibnamefont{and}
  \bibinfo{author}{\bibfnamefont{K.}~\bibnamefont{Kuroki}},
  \bibinfo{journal}{arXiv:1909.00060}  (\bibinfo{year}{2019}).

\bibitem[{\citenamefont{Gao et~al.}(2019)\citenamefont{Gao, Wang, Fang, and
  Weng}}]{Gao2019}
\bibinfo{author}{\bibfnamefont{J.}~\bibnamefont{Gao}},
  \bibinfo{author}{\bibfnamefont{Z.}~\bibnamefont{Wang}},
  \bibinfo{author}{\bibfnamefont{C.}~\bibnamefont{Fang}}, \bibnamefont{and}
  \bibinfo{author}{\bibfnamefont{H.}~\bibnamefont{Weng}},
  \bibinfo{journal}{arXiv:1909.04657}  (\bibinfo{year}{2019}).

\bibitem[{\citenamefont{{Bernardini} et~al.}(2020)\citenamefont{{Bernardini},
  {Olevano}, and {Cano}}}]{bernardini2020magnetic}
\bibinfo{author}{\bibfnamefont{F.}~\bibnamefont{{Bernardini}}},
  \bibinfo{author}{\bibfnamefont{V.}~\bibnamefont{{Olevano}}},
  \bibnamefont{and} \bibinfo{author}{\bibfnamefont{A.}~\bibnamefont{{Cano}}},
  \bibinfo{journal}{Phys. Rev. Ressearch} \textbf{\bibinfo{volume}{2}},
  \bibinfo{pages}{013219} (\bibinfo{year}{2020}).

\bibitem[{\citenamefont{Jiang et~al.}(2019)\citenamefont{Jiang, Si, Liao, and
  Zhong}}]{Jiang2019electronic}
\bibinfo{author}{\bibfnamefont{P.}~\bibnamefont{Jiang}},
  \bibinfo{author}{\bibfnamefont{L.}~\bibnamefont{Si}},
  \bibinfo{author}{\bibfnamefont{Z.}~\bibnamefont{Liao}}, \bibnamefont{and}
  \bibinfo{author}{\bibfnamefont{Z.}~\bibnamefont{Zhong}},
  \bibinfo{journal}{Phys. Rev. B} \textbf{\bibinfo{volume}{100}},
  \bibinfo{pages}{201106(R)} (\bibinfo{year}{2019}).

\bibitem[{\citenamefont{Hayward and Rosseinsky}(2003)}]{hayward2003}
\bibinfo{author}{\bibfnamefont{M.}~\bibnamefont{Hayward}} \bibnamefont{and}
  \bibinfo{author}{\bibfnamefont{M.}~\bibnamefont{Rosseinsky}},
  \bibinfo{journal}{Solid State Sci.} \textbf{\bibinfo{volume}{5}},
  \bibinfo{pages}{839} (\bibinfo{year}{2003}).

\bibitem[{\citenamefont{Zhang et~al.}(2020{\natexlab{a}})\citenamefont{Zhang,
  Jin, Wang, Xi, Shi, Ye, and Mei}}]{HZhang2019}
\bibinfo{author}{\bibfnamefont{H.}~\bibnamefont{Zhang}},
  \bibinfo{author}{\bibfnamefont{L.}~\bibnamefont{Jin}},
  \bibinfo{author}{\bibfnamefont{S.}~\bibnamefont{Wang}},
  \bibinfo{author}{\bibfnamefont{B.}~\bibnamefont{Xi}},
  \bibinfo{author}{\bibfnamefont{X.}~\bibnamefont{Shi}},
  \bibinfo{author}{\bibfnamefont{F.}~\bibnamefont{Ye}}, \bibnamefont{and}
  \bibinfo{author}{\bibfnamefont{J.-W.} \bibnamefont{Mei}},
  \bibinfo{journal}{Phys. Rev. Research} \textbf{\bibinfo{volume}{2}},
  \bibinfo{pages}{013214} (\bibinfo{year}{2020}{\natexlab{a}}).

\bibitem[{\citenamefont{Wu et~al.}(2020)\citenamefont{Wu, Sante, Schwemmer,
  Hanke, Hwang, Raghu, and Thomale}}]{Wu2019}
\bibinfo{author}{\bibfnamefont{X.}~\bibnamefont{Wu}},
  \bibinfo{author}{\bibfnamefont{D.~D.} \bibnamefont{Sante}},
  \bibinfo{author}{\bibfnamefont{T.}~\bibnamefont{Schwemmer}},
  \bibinfo{author}{\bibfnamefont{W.}~\bibnamefont{Hanke}},
  \bibinfo{author}{\bibfnamefont{H.~Y.} \bibnamefont{Hwang}},
  \bibinfo{author}{\bibfnamefont{S.}~\bibnamefont{Raghu}}, \bibnamefont{and}
  \bibinfo{author}{\bibfnamefont{R.}~\bibnamefont{Thomale}},
  \bibinfo{journal}{Phys. Rev. B} \textbf{\bibinfo{volume}{101}},
  \bibinfo{pages}{060504} (\bibinfo{year}{2020}).

\bibitem[{\citenamefont{Nomura et~al.}(2019)\citenamefont{Nomura, Hirayama,
  Tadano, Yoshimoto, Nakamura, and Arita}}]{Normura2019}
\bibinfo{author}{\bibfnamefont{Y.}~\bibnamefont{Nomura}},
  \bibinfo{author}{\bibfnamefont{M.}~\bibnamefont{Hirayama}},
  \bibinfo{author}{\bibfnamefont{T.}~\bibnamefont{Tadano}},
  \bibinfo{author}{\bibfnamefont{Y.}~\bibnamefont{Yoshimoto}},
  \bibinfo{author}{\bibfnamefont{K.}~\bibnamefont{Nakamura}}, \bibnamefont{and}
  \bibinfo{author}{\bibfnamefont{R.}~\bibnamefont{Arita}},
  \bibinfo{journal}{Phys. Rev. B} \textbf{\bibinfo{volume}{100}},
  \bibinfo{pages}{205138} (\bibinfo{year}{2019}).

\bibitem[{\citenamefont{Singh}(2019)}]{Singh2019}
\bibinfo{author}{\bibfnamefont{N.}~\bibnamefont{Singh}},
  \bibinfo{journal}{arXiv:1909.07688}  (\bibinfo{year}{2019}).

\bibitem[{\citenamefont{{Choi} et~al.}(2020)\citenamefont{{Choi}, {Lee}, and
  {Pickett}}}]{choi2020role}
\bibinfo{author}{\bibfnamefont{M.-Y.} \bibnamefont{{Choi}}},
  \bibinfo{author}{\bibfnamefont{K.-W.} \bibnamefont{{Lee}}}, \bibnamefont{and}
  \bibinfo{author}{\bibfnamefont{W.~E.} \bibnamefont{{Pickett}}},
  \bibinfo{journal}{Phys. Rev. B} \textbf{\bibinfo{volume}{101}},
  \bibinfo{pages}{020503} (\bibinfo{year}{2020}).

\bibitem[{\citenamefont{Jiang et~al.}(2020)\citenamefont{Jiang, Berciu, and
  Sawatzky}}]{Jiang2019}
\bibinfo{author}{\bibfnamefont{M.}~\bibnamefont{Jiang}},
  \bibinfo{author}{\bibfnamefont{M.}~\bibnamefont{Berciu}}, \bibnamefont{and}
  \bibinfo{author}{\bibfnamefont{G.~A.} \bibnamefont{Sawatzky}},
  \bibinfo{journal}{Phys. Rev. Lett.} \textbf{\bibinfo{volume}{124}},
  \bibinfo{pages}{207004} (\bibinfo{year}{2020}).

\bibitem[{\citenamefont{Zhang and Vishwanath}(2019)}]{YHZhang2019}
\bibinfo{author}{\bibfnamefont{Y.-H.} \bibnamefont{Zhang}} \bibnamefont{and}
  \bibinfo{author}{\bibfnamefont{A.}~\bibnamefont{Vishwanath}},
  \bibinfo{journal}{arXiv:1909.12865}  (\bibinfo{year}{2019}).

\bibitem[{\citenamefont{Hu and Wu}(2019)}]{Hu2019}
\bibinfo{author}{\bibfnamefont{L.-H.} \bibnamefont{Hu}} \bibnamefont{and}
  \bibinfo{author}{\bibfnamefont{C.}~\bibnamefont{Wu}}, \bibinfo{journal}{Phys.
  Rev. Research} \textbf{\bibinfo{volume}{1}}, \bibinfo{pages}{032046}
  (\bibinfo{year}{2019}).

\bibitem[{\citenamefont{Werner and Hoshino}(2020)}]{Werner2019}
\bibinfo{author}{\bibfnamefont{P.}~\bibnamefont{Werner}} \bibnamefont{and}
  \bibinfo{author}{\bibfnamefont{S.}~\bibnamefont{Hoshino}},
  \bibinfo{journal}{Phys. Rev. B} \textbf{\bibinfo{volume}{101}},
  \bibinfo{pages}{041104} (\bibinfo{year}{2020}).

\bibitem[{\citenamefont{Zhang et~al.}(2020{\natexlab{b}})\citenamefont{Zhang,
  Yang, and Zhang}}]{GMZhang2019}
\bibinfo{author}{\bibfnamefont{G.-M.} \bibnamefont{Zhang}},
  \bibinfo{author}{\bibfnamefont{Y.-F.} \bibnamefont{Yang}}, \bibnamefont{and}
  \bibinfo{author}{\bibfnamefont{F.-C.} \bibnamefont{Zhang}},
  \bibinfo{journal}{Phys. Rev. B} \textbf{\bibinfo{volume}{101}},
  \bibinfo{pages}{020501} (\bibinfo{year}{2020}{\natexlab{b}}).

\bibitem[{\citenamefont{Fu et~al.}(2019)\citenamefont{Fu, Wang, Cheng, Pei,
  Zhou, Chen, Wang, Zhao, Jiang, Liu et~al.}}]{fu2019corelevel}
\bibinfo{author}{\bibfnamefont{Y.}~\bibnamefont{Fu}},
  \bibinfo{author}{\bibfnamefont{L.}~\bibnamefont{Wang}},
  \bibinfo{author}{\bibfnamefont{H.}~\bibnamefont{Cheng}},
  \bibinfo{author}{\bibfnamefont{S.}~\bibnamefont{Pei}},
  \bibinfo{author}{\bibfnamefont{X.}~\bibnamefont{Zhou}},
  \bibinfo{author}{\bibfnamefont{J.}~\bibnamefont{Chen}},
  \bibinfo{author}{\bibfnamefont{S.}~\bibnamefont{Wang}},
  \bibinfo{author}{\bibfnamefont{R.}~\bibnamefont{Zhao}},
  \bibinfo{author}{\bibfnamefont{W.}~\bibnamefont{Jiang}},
  \bibinfo{author}{\bibfnamefont{C.}~\bibnamefont{Liu}}, \bibnamefont{et~al.},
  \bibinfo{journal}{arXiv:1911.03177}  (\bibinfo{year}{2019}).

\bibitem[{\citenamefont{Li et~al.}(2020{\natexlab{a}})\citenamefont{Li, He, Si,
  Zhu, Zhang, and Wen}}]{qLi2019}
\bibinfo{author}{\bibfnamefont{Q.}~\bibnamefont{Li}},
  \bibinfo{author}{\bibfnamefont{C.}~\bibnamefont{He}},
  \bibinfo{author}{\bibfnamefont{J.}~\bibnamefont{Si}},
  \bibinfo{author}{\bibfnamefont{X.}~\bibnamefont{Zhu}},
  \bibinfo{author}{\bibfnamefont{Y.}~\bibnamefont{Zhang}}, \bibnamefont{and}
  \bibinfo{author}{\bibfnamefont{H.-H.} \bibnamefont{Wen}},
  \bibinfo{journal}{Nat. Commun.} \textbf{\bibinfo{volume}{1}},
  \bibinfo{pages}{1} (\bibinfo{year}{2020}{\natexlab{a}}).

\bibitem[{\citenamefont{Zhou et~al.}(2020)\citenamefont{Zhou, Feng, Qin, Yan,
  Hu, Guo, Wang, Wu, Zhang, Chen et~al.}}]{xZhou2019}
\bibinfo{author}{\bibfnamefont{X.}~\bibnamefont{Zhou}},
  \bibinfo{author}{\bibfnamefont{Z.}~\bibnamefont{Feng}},
  \bibinfo{author}{\bibfnamefont{P.}~\bibnamefont{Qin}},
  \bibinfo{author}{\bibfnamefont{H.}~\bibnamefont{Yan}},
  \bibinfo{author}{\bibfnamefont{S.}~\bibnamefont{Hu}},
  \bibinfo{author}{\bibfnamefont{H.}~\bibnamefont{Guo}},
  \bibinfo{author}{\bibfnamefont{X.}~\bibnamefont{Wang}},
  \bibinfo{author}{\bibfnamefont{H.}~\bibnamefont{Wu}},
  \bibinfo{author}{\bibfnamefont{X.}~\bibnamefont{Zhang}},
  \bibinfo{author}{\bibfnamefont{H.}~\bibnamefont{Chen}}, \bibnamefont{et~al.},
  \bibinfo{journal}{Rare Met.} \textbf{\bibinfo{volume}{39}},
  \bibinfo{pages}{368} (\bibinfo{year}{2020}).

\bibitem[{\citenamefont{Si et~al.}(2019)\citenamefont{Si, Xiao, Kaufmann,
  Tomczak, Lu, Zhong, and Held}}]{si2019topotactic}
\bibinfo{author}{\bibfnamefont{L.}~\bibnamefont{Si}},
  \bibinfo{author}{\bibfnamefont{W.}~\bibnamefont{Xiao}},
  \bibinfo{author}{\bibfnamefont{J.}~\bibnamefont{Kaufmann}},
  \bibinfo{author}{\bibfnamefont{J.~M.} \bibnamefont{Tomczak}},
  \bibinfo{author}{\bibfnamefont{Y.}~\bibnamefont{Lu}},
  \bibinfo{author}{\bibfnamefont{Z.}~\bibnamefont{Zhong}}, \bibnamefont{and}
  \bibinfo{author}{\bibfnamefont{K.}~\bibnamefont{Held}},
  \bibinfo{journal}{arXiv:1911.06917}  (\bibinfo{year}{2019}).

\bibitem[{\citenamefont{Lechermann}(2020)}]{lechermann2019late}
\bibinfo{author}{\bibfnamefont{F.}~\bibnamefont{Lechermann}},
  \bibinfo{journal}{Phys. Rev. B} \textbf{\bibinfo{volume}{101}},
  \bibinfo{pages}{081110} (\bibinfo{year}{2020}).

\bibitem[{\citenamefont{Lee et~al.}(2018)\citenamefont{Lee, Chubukov, Miao, and
  Kotliar}}]{Lee2018}
\bibinfo{author}{\bibfnamefont{T.-H.} \bibnamefont{Lee}},
  \bibinfo{author}{\bibfnamefont{A.}~\bibnamefont{Chubukov}},
  \bibinfo{author}{\bibfnamefont{H.}~\bibnamefont{Miao}}, \bibnamefont{and}
  \bibinfo{author}{\bibfnamefont{G.}~\bibnamefont{Kotliar}},
  \bibinfo{journal}{Phys. Rev. Lett.} \textbf{\bibinfo{volume}{121}},
  \bibinfo{pages}{187003} (\bibinfo{year}{2018}).

\bibitem[{\citenamefont{Georges et~al.}(2013)\citenamefont{Georges, Medici, and
  Mravlje}}]{Georges2013}
\bibinfo{author}{\bibfnamefont{A.}~\bibnamefont{Georges}},
  \bibinfo{author}{\bibfnamefont{L.~d.} \bibnamefont{Medici}},
  \bibnamefont{and} \bibinfo{author}{\bibfnamefont{J.}~\bibnamefont{Mravlje}},
  \bibinfo{journal}{Annual Review of Condensed Matter Physics}
  \textbf{\bibinfo{volume}{4}}, \bibinfo{pages}{137} (\bibinfo{year}{2013}).

\bibitem[{\citenamefont{Haule and Kotliar}(2009)}]{Haule2009}
\bibinfo{author}{\bibfnamefont{K.}~\bibnamefont{Haule}} \bibnamefont{and}
  \bibinfo{author}{\bibfnamefont{G.}~\bibnamefont{Kotliar}},
  \bibinfo{journal}{New Journal of Physics} \textbf{\bibinfo{volume}{11}},
  \bibinfo{pages}{025021} (\bibinfo{year}{2009}).

\bibitem[{\citenamefont{Werner et~al.}(2008)\citenamefont{Werner, Gull, Troyer,
  and Millis}}]{Werner2008}
\bibinfo{author}{\bibfnamefont{P.}~\bibnamefont{Werner}},
  \bibinfo{author}{\bibfnamefont{E.}~\bibnamefont{Gull}},
  \bibinfo{author}{\bibfnamefont{M.}~\bibnamefont{Troyer}}, \bibnamefont{and}
  \bibinfo{author}{\bibfnamefont{A.~J.} \bibnamefont{Millis}},
  \bibinfo{journal}{Phys. Rev. Lett.} \textbf{\bibinfo{volume}{101}},
  \bibinfo{pages}{166405} (\bibinfo{year}{2008}).

\bibitem[{\citenamefont{Chang and Zhao}(2017)}]{Chang2017}
\bibinfo{author}{\bibfnamefont{J.}~\bibnamefont{Chang}} \bibnamefont{and}
  \bibinfo{author}{\bibfnamefont{J.}~\bibnamefont{Zhao}},
  \bibinfo{journal}{Eur. Phys. J. B} \textbf{\bibinfo{volume}{90}},
  \bibinfo{pages}{154} (\bibinfo{year}{2017}).

\bibitem[{\citenamefont{Zaanen and Ole\ifmmode~\acute{s}\else
  \'{s}\fi{}}(1988)}]{Zaanen1988}
\bibinfo{author}{\bibfnamefont{J.}~\bibnamefont{Zaanen}} \bibnamefont{and}
  \bibinfo{author}{\bibfnamefont{A.~M.} \bibnamefont{Ole\ifmmode~\acute{s}\else
  \'{s}\fi{}}}, \bibinfo{journal}{Phys. Rev. B} \textbf{\bibinfo{volume}{37}},
  \bibinfo{pages}{9423} (\bibinfo{year}{1988}).

\bibitem[{\citenamefont{Hayden et~al.}(1991)\citenamefont{Hayden, Aeppli, Mook,
  Rytz, Hundley, and Fisk}}]{Hayden1991}
\bibinfo{author}{\bibfnamefont{S.~M.} \bibnamefont{Hayden}},
  \bibinfo{author}{\bibfnamefont{G.}~\bibnamefont{Aeppli}},
  \bibinfo{author}{\bibfnamefont{H.}~\bibnamefont{Mook}},
  \bibinfo{author}{\bibfnamefont{D.}~\bibnamefont{Rytz}},
  \bibinfo{author}{\bibfnamefont{M.~F.} \bibnamefont{Hundley}},
  \bibnamefont{and} \bibinfo{author}{\bibfnamefont{Z.}~\bibnamefont{Fisk}},
  \bibinfo{journal}{Phys. Rev. Lett.} \textbf{\bibinfo{volume}{66}},
  \bibinfo{pages}{821} (\bibinfo{year}{1991}).

\bibitem[{\citenamefont{Keimer et~al.}(1991)\citenamefont{Keimer, Birgeneau,
  Cassanho, Endoh, Erwin, Kastner, and Shirane}}]{Keimer1991}
\bibinfo{author}{\bibfnamefont{B.}~\bibnamefont{Keimer}},
  \bibinfo{author}{\bibfnamefont{R.~J.} \bibnamefont{Birgeneau}},
  \bibinfo{author}{\bibfnamefont{A.}~\bibnamefont{Cassanho}},
  \bibinfo{author}{\bibfnamefont{Y.}~\bibnamefont{Endoh}},
  \bibinfo{author}{\bibfnamefont{R.~W.} \bibnamefont{Erwin}},
  \bibinfo{author}{\bibfnamefont{M.~A.} \bibnamefont{Kastner}},
  \bibnamefont{and} \bibinfo{author}{\bibfnamefont{G.}~\bibnamefont{Shirane}},
  \bibinfo{journal}{Phys. Rev. Lett.} \textbf{\bibinfo{volume}{67}},
  \bibinfo{pages}{1930} (\bibinfo{year}{1991}).

\bibitem[{\citenamefont{Tranquada et~al.}(1992)\citenamefont{Tranquada,
  Gehring, Shirane, Shamoto, and Sato}}]{Tranquada1992}
\bibinfo{author}{\bibfnamefont{J.~M.} \bibnamefont{Tranquada}},
  \bibinfo{author}{\bibfnamefont{P.~M.} \bibnamefont{Gehring}},
  \bibinfo{author}{\bibfnamefont{G.}~\bibnamefont{Shirane}},
  \bibinfo{author}{\bibfnamefont{S.}~\bibnamefont{Shamoto}}, \bibnamefont{and}
  \bibinfo{author}{\bibfnamefont{M.}~\bibnamefont{Sato}},
  \bibinfo{journal}{Phys. Rev. B} \textbf{\bibinfo{volume}{46}},
  \bibinfo{pages}{5561} (\bibinfo{year}{1992}).

\bibitem[{\citenamefont{Zhang and Rice}(1988)}]{Zhang1988}
\bibinfo{author}{\bibfnamefont{F.~C.} \bibnamefont{Zhang}} \bibnamefont{and}
  \bibinfo{author}{\bibfnamefont{T.~M.} \bibnamefont{Rice}},
  \bibinfo{journal}{Phys. Rev. B} \textbf{\bibinfo{volume}{37}},
  \bibinfo{pages}{3759} (\bibinfo{year}{1988}).

\bibitem[{\citenamefont{Coleman}(2007)}]{Coleman2007}
\bibinfo{author}{\bibfnamefont{P.}~\bibnamefont{Coleman}}, in
  \emph{\bibinfo{booktitle}{Handbook of Magnetism and Advanced Magnetic
  Materials}}, edited by
  \bibinfo{editor}{\bibfnamefont{H.}~\bibnamefont{Kronm\"uller}}
  \bibnamefont{and} \bibinfo{editor}{\bibfnamefont{S.}~\bibnamefont{Parkin}}
  (\bibinfo{publisher}{John Wiley and Sons, New York}, \bibinfo{year}{2007}),
  vol.~\bibinfo{volume}{1}, pp. \bibinfo{pages}{95--148}.

\bibitem[{\citenamefont{Scalapino et~al.}(1986)\citenamefont{Scalapino, Loh,
  and Hirsch}}]{Scalapino1986}
\bibinfo{author}{\bibfnamefont{D.~J.} \bibnamefont{Scalapino}},
  \bibinfo{author}{\bibfnamefont{E.}~\bibnamefont{Loh}}, \bibnamefont{and}
  \bibinfo{author}{\bibfnamefont{J.~E.} \bibnamefont{Hirsch}},
  \bibinfo{journal}{Phys. Rev. B} \textbf{\bibinfo{volume}{34}},
  \bibinfo{pages}{8190} (\bibinfo{year}{1986}).

\bibitem[{\citenamefont{Bickers et~al.}(1987)\citenamefont{Bickers, Scalapino,
  and Scalettar}}]{Bickers1987}
\bibinfo{author}{\bibfnamefont{N.}~\bibnamefont{Bickers}},
  \bibinfo{author}{\bibfnamefont{D.}~\bibnamefont{Scalapino}},
  \bibnamefont{and}
  \bibinfo{author}{\bibfnamefont{R.}~\bibnamefont{Scalettar}},
  \bibinfo{journal}{Int. J. Mod. Phys. B} \textbf{\bibinfo{volume}{01}},
  \bibinfo{pages}{687} (\bibinfo{year}{1987}).

\bibitem[{\citenamefont{Inui et~al.}(1988)\citenamefont{Inui, Doniach,
  Hirschfeld, and Ruckenstein}}]{Inui1988}
\bibinfo{author}{\bibfnamefont{M.}~\bibnamefont{Inui}},
  \bibinfo{author}{\bibfnamefont{S.}~\bibnamefont{Doniach}},
  \bibinfo{author}{\bibfnamefont{P.~J.} \bibnamefont{Hirschfeld}},
  \bibnamefont{and} \bibinfo{author}{\bibfnamefont{A.~E.}
  \bibnamefont{Ruckenstein}}, \bibinfo{journal}{Phys. Rev. B}
  \textbf{\bibinfo{volume}{37}}, \bibinfo{pages}{2320} (\bibinfo{year}{1988}).

\bibitem[{\citenamefont{Dong et~al.}(1988)\citenamefont{Dong, Liang, Che, Xie,
  Zhao, Yang, Ni, and Liu}}]{Dong1988}
\bibinfo{author}{\bibfnamefont{C.}~\bibnamefont{Dong}},
  \bibinfo{author}{\bibfnamefont{J.~K.} \bibnamefont{Liang}},
  \bibinfo{author}{\bibfnamefont{G.~C.} \bibnamefont{Che}},
  \bibinfo{author}{\bibfnamefont{S.~S.} \bibnamefont{Xie}},
  \bibinfo{author}{\bibfnamefont{Z.~X.} \bibnamefont{Zhao}},
  \bibinfo{author}{\bibfnamefont{Q.~S.} \bibnamefont{Yang}},
  \bibinfo{author}{\bibfnamefont{Y.~M.} \bibnamefont{Ni}}, \bibnamefont{and}
  \bibinfo{author}{\bibfnamefont{G.~R.} \bibnamefont{Liu}},
  \bibinfo{journal}{Phys. Rev. B} \textbf{\bibinfo{volume}{37}},
  \bibinfo{pages}{5182} (\bibinfo{year}{1988}).

\bibitem[{\citenamefont{Kotliar and Liu}(1988)}]{Kotliar1988}
\bibinfo{author}{\bibfnamefont{G.}~\bibnamefont{Kotliar}} \bibnamefont{and}
  \bibinfo{author}{\bibfnamefont{J.}~\bibnamefont{Liu}},
  \bibinfo{journal}{Phys. Rev. B} \textbf{\bibinfo{volume}{38}},
  \bibinfo{pages}{5142} (\bibinfo{year}{1988}).

\bibitem[{\citenamefont{Monthoux et~al.}(1991)\citenamefont{Monthoux, Balatsky,
  and Pines}}]{Monthoux1991}
\bibinfo{author}{\bibfnamefont{P.}~\bibnamefont{Monthoux}},
  \bibinfo{author}{\bibfnamefont{A.~V.} \bibnamefont{Balatsky}},
  \bibnamefont{and} \bibinfo{author}{\bibfnamefont{D.}~\bibnamefont{Pines}},
  \bibinfo{journal}{Phys. Rev. Lett.} \textbf{\bibinfo{volume}{67}},
  \bibinfo{pages}{3448} (\bibinfo{year}{1991}).

\bibitem[{\citenamefont{Moriya et~al.}(1990)\citenamefont{Moriya, Takahashi,
  and Ueda}}]{Moriya1990}
\bibinfo{author}{\bibfnamefont{T.}~\bibnamefont{Moriya}},
  \bibinfo{author}{\bibfnamefont{Y.}~\bibnamefont{Takahashi}},
  \bibnamefont{and} \bibinfo{author}{\bibfnamefont{K.}~\bibnamefont{Ueda}},
  \bibinfo{journal}{J. Phys. Soc. Jpn.} \textbf{\bibinfo{volume}{59}},
  \bibinfo{pages}{2905} (\bibinfo{year}{1990}).

\bibitem[{\citenamefont{Millis et~al.}(1990)\citenamefont{Millis, Monien, and
  Pines}}]{Millis1990}
\bibinfo{author}{\bibfnamefont{A.~J.} \bibnamefont{Millis}},
  \bibinfo{author}{\bibfnamefont{H.}~\bibnamefont{Monien}}, \bibnamefont{and}
  \bibinfo{author}{\bibfnamefont{D.}~\bibnamefont{Pines}},
  \bibinfo{journal}{Phys. Rev. B} \textbf{\bibinfo{volume}{42}},
  \bibinfo{pages}{167} (\bibinfo{year}{1990}).

\bibitem[{\citenamefont{Chubukov et~al.}(2008)\citenamefont{Chubukov, Pines,
  and Schmalian}}]{Chubukov2008}
\bibinfo{author}{\bibfnamefont{A.}~\bibnamefont{Chubukov}},
  \bibinfo{author}{\bibfnamefont{D.}~\bibnamefont{Pines}}, \bibnamefont{and}
  \bibinfo{author}{\bibfnamefont{J.}~\bibnamefont{Schmalian}}, in
  \emph{\bibinfo{booktitle}{Superconductivity}}, edited by
  \bibinfo{editor}{\bibfnamefont{K.}~\bibnamefont{Bennemann}} \bibnamefont{and}
  \bibinfo{editor}{\bibfnamefont{J.}~\bibnamefont{Ketterson}}
  (\bibinfo{publisher}{Springer Berlin Heidelberg}, \bibinfo{year}{2008}), pp.
  \bibinfo{pages}{1349--1413}.

\bibitem[{\citenamefont{Scalapino}(2012)}]{Scalapino2012}
\bibinfo{author}{\bibfnamefont{D.~J.} \bibnamefont{Scalapino}},
  \bibinfo{journal}{Rev. Mod. Phys.} \textbf{\bibinfo{volume}{84}},
  \bibinfo{pages}{1383} (\bibinfo{year}{2012}).

\bibitem[{\citenamefont{Böker et~al.}(2020)\citenamefont{Böker, Sulangi,
  Akbari, Davis, Hirschfeld, and Eremin}}]{bker2020phasesensitive}
\bibinfo{author}{\bibfnamefont{J.}~\bibnamefont{Böker}},
  \bibinfo{author}{\bibfnamefont{M.~A.} \bibnamefont{Sulangi}},
  \bibinfo{author}{\bibfnamefont{A.}~\bibnamefont{Akbari}},
  \bibinfo{author}{\bibfnamefont{J.~C.~S.} \bibnamefont{Davis}},
  \bibinfo{author}{\bibfnamefont{P.~J.} \bibnamefont{Hirschfeld}},
  \bibnamefont{and} \bibinfo{author}{\bibfnamefont{I.~M.}
  \bibnamefont{Eremin}}, \bibinfo{journal}{arXiv:2004.02768}
  (\bibinfo{year}{2020}).

\bibitem[{\citenamefont{Gu et~al.}(2019)\citenamefont{Gu, Zhu, Wang, Hu, and
  Chen}}]{gu2019hybridization}
\bibinfo{author}{\bibfnamefont{Y.}~\bibnamefont{Gu}},
  \bibinfo{author}{\bibfnamefont{S.}~\bibnamefont{Zhu}},
  \bibinfo{author}{\bibfnamefont{X.}~\bibnamefont{Wang}},
  \bibinfo{author}{\bibfnamefont{J.}~\bibnamefont{Hu}}, \bibnamefont{and}
  \bibinfo{author}{\bibfnamefont{H.}~\bibnamefont{Chen}},
  \bibinfo{journal}{arXiv:1911.00814}  (\bibinfo{year}{2019}).

\bibitem[{\citenamefont{Li et~al.}(2020{\natexlab{b}})\citenamefont{Li, Wang,
  Lee, Harvey, Osada, Goodge, Kourkoutis, and Hwang}}]{li2020superconducting}
\bibinfo{author}{\bibfnamefont{D.}~\bibnamefont{Li}},
  \bibinfo{author}{\bibfnamefont{B.~Y.} \bibnamefont{Wang}},
  \bibinfo{author}{\bibfnamefont{K.}~\bibnamefont{Lee}},
  \bibinfo{author}{\bibfnamefont{S.~P.} \bibnamefont{Harvey}},
  \bibinfo{author}{\bibfnamefont{M.}~\bibnamefont{Osada}},
  \bibinfo{author}{\bibfnamefont{B.~H.} \bibnamefont{Goodge}},
  \bibinfo{author}{\bibfnamefont{L.~F.} \bibnamefont{Kourkoutis}},
  \bibnamefont{and} \bibinfo{author}{\bibfnamefont{H.~Y.} \bibnamefont{Hwang}},
  \bibinfo{journal}{arXiv:2003.08506}  (\bibinfo{year}{2020}{\natexlab{b}}).

\bibitem[{\citenamefont{Kaindl et~al.}(1989)\citenamefont{Kaindl, Strebel,
  Kolodziejczyk, Sch{\"a}fer, Kiemel, L{\"o}sch, Kemmler-Sack, Hoppe,
  M{\"u}ller, and Kissel}}]{kaindl1989correlation}
\bibinfo{author}{\bibfnamefont{G.}~\bibnamefont{Kaindl}},
  \bibinfo{author}{\bibfnamefont{O.}~\bibnamefont{Strebel}},
  \bibinfo{author}{\bibfnamefont{A.}~\bibnamefont{Kolodziejczyk}},
  \bibinfo{author}{\bibfnamefont{W.}~\bibnamefont{Sch{\"a}fer}},
  \bibinfo{author}{\bibfnamefont{R.}~\bibnamefont{Kiemel}},
  \bibinfo{author}{\bibfnamefont{S.}~\bibnamefont{L{\"o}sch}},
  \bibinfo{author}{\bibfnamefont{S.}~\bibnamefont{Kemmler-Sack}},
  \bibinfo{author}{\bibfnamefont{R.}~\bibnamefont{Hoppe}},
  \bibinfo{author}{\bibfnamefont{H.}~\bibnamefont{M{\"u}ller}},
  \bibnamefont{and} \bibinfo{author}{\bibfnamefont{D.}~\bibnamefont{Kissel}},
  \bibinfo{journal}{Physica B: Condensed Matter}
  \textbf{\bibinfo{volume}{158}}, \bibinfo{pages}{446} (\bibinfo{year}{1989}).

\end{thebibliography}
\end{document}